\documentstyle[11pt,twoside,epsfig]{article}
\newcommand{\mb}[1]{\mbox{\scriptsize\it #1}}

\newlength{\dinwidth}
\newlength{\dinmargin}
\newcommand{\lsim}{\raisebox{-0.5mm}{$\stackrel{<}{\scriptstyle{\sim}}$}}
\newcommand{\isge }{\sigma(E)/E}
\newcommand{\comm}[1]{\mbox{#1}}
\newcommand{\rb}[1]{\raisebox{1.5ex}[-2.5ex]{#1}}
\newcommand{\va   }{\vphantom{$\displaystyle{a\over b}$}}
\newcommand{\mref }{M_{\it ref}}
\newcommand{\zn   }{Z^{\circ}}
\newcommand{\al   }{\alpha_{s}}
\newcommand{\alq  }{\al(\qq)}
\newcommand{\alsq }{\al^2}
\newcommand{\mzq  }{M_Z^{2}}
\newcommand{\almz }{\al(\mzq)}
\newcommand{\almu }{\al(\murq)}
\newcommand{\alcena}{0.117 \pm 0.003 \,(stat)}
\newcommand{\almeana}{0.117 \pm 0.003 \,(stat) \, 
                            ^{+\,0.009}_{-\,0.013} \,(sys)}
\newcommand{\almeanars}{0.117 \pm 0.003 \,(stat) \, 
                              ^{+\,0.009}_{-\,0.013} \,(sys) \,
                              + 0.006 \,(jet\; algorithm)}
\newcommand{\msb  }{\scriptstyle \overline{\it{MS}}}
\newcommand{\lav  }{\Lambda^{(4)}_{\msb}}
\newcommand{\oas  }{{\cal O}(\al)}
\newcommand{\oasd }{{\cal O}(\alsq)}
\newcommand{\qq   }{Q^{2}}
\newcommand{\mij  }{m^{2}_{ij}}
\newcommand{\gev  }{GeV}
\newcommand{\gevq }{GeV$^2\,$}
\newcommand{\tj   }{\theta_{jet}}
\newcommand{\murq }{\mu^{2}_r}

\begin{document}
\begin{titlepage}
\begin{flushleft}
%
%
{\tt DESY 98-087  \hfill ISSN 0418-9833 \\}
\end{flushleft}
\vspace*{4.cm}
\begin{center}
\begin{Large}
%
%
{\bf Multi-Jet Event Rates in Deep Inelastic Scattering \\ 
     and Determination of the Strong Coupling Constant} \\
\vspace*{2.cm}
H1 Collaboration \\
\end{Large}
\vspace*{4.cm}
{\bf Abstract:}
\begin{quotation}
%
%
%
Jet event rates in deep inelastic {\it ep}$\,$ scattering at HERA are 
investigated applying the modified JADE jet algorithm. The data are 
corrected for detector and hadronization effects and then compared 
with perturbative QCD predictions using next-to-leading order calculations. 
The strong coupling constant $\almz$ is determined evaluating the jet event 
rates. Values of $\alq$ are extracted in four different bins of the negative 
squared momentum transfer~$\qq$ in the range from 40~\gevq to 4000~\gevq. A 
combined fit of the renormalization group equation to these several 
$\alq$ values results in $\almz = \almeanars$. 

\end{quotation}
\vspace*{4cm}
%
%
{\it Submitted to Eur.~Phys.~J.}
\cleardoublepage
\end{center}
\end{titlepage}
\begin{flushleft}
 C.~Adloff$^{34}$,                
 M.~Anderson$^{22}$,              
 V.~Andreev$^{25}$,               
 B.~Andrieu$^{28}$,               
 V.~Arkadov$^{35}$,               
 C.~Arndt$^{11}$,                 
 I.~Ayyaz$^{29}$,                 
 A.~Babaev$^{24}$,                
 J.~B\"ahr$^{35}$,                
 J.~B\'an$^{17}$
% ...

\end{flushleft}
\begin{flushleft} 
   {\it 
% ...
$ ^{11}$ DESY, Hamburg, Germany$^a$ \\
% ...
$ ^{17}$ Institute of Experimental Physics, Slovak Academy of
          Sciences, Ko\v{s}ice, Slovak Republic$^{f,j}$ \\
% ...
$ ^{22}$ Department of Physics and Astronomy, 
          University of Manchester, Manchester, UK$^b$ \\
% ...
$ ^{24}$ Institute for Theoretical and Experimental Physics,
          Moscow, Russia \\
 $ ^{25}$ Lebedev Physical Institute, Moscow, Russia$^{f,k}$ \\
% ...
$ ^{28}$ LPNHE, Ecole Polytechnique, IN2P3-CNRS, Palaiseau, France \\
 $ ^{29}$ LPNHE, Universit\'{e}s Paris VI and VII, IN2P3-CNRS,
          Paris, France \\
% ...
$ ^{34}$ Fachbereich Physik, Bergische Universit\"at Gesamthochschule
          Wuppertal, Wuppertal, Germany$^a$ \\
 $ ^{35}$ DESY, Institut f\"ur Hochenergiephysik, Zeuthen, Germany$^a$ \\
% ...
$ ^a$ Supported by the Bundesministerium f\"ur Bildung, Wissenschaft,
        Forschung und Technologie, FRG,
        under contract numbers 7AC17P, 7AC47P, 7DO55P, 7HH17I, 7HH27P,
        7HD17P, 7HD27P, 7KI17I, 6MP17I and 7WT87P \\
 $ ^b$ Supported by the UK Particle Physics and Astronomy Research
       Council, and formerly by the UK Science and Engineering Research
       Council \\
% ...
$ ^f$ Supported by the Deutsche Forschungsgemeinschaft \\
% ...
$ ^j$ Supported by VEGA SR grant no. 2/5167/98 \\
 $ ^k$ Supported by Russian Foundation for Basic Research 
       grant no. 96-02-00019 
 } 
\end{flushleft}
%
\newpage
%
%
%
\section{Introduction}
\label{intro}
Multi-jet production in neutral current deep
inelastic positron proton scattering (DIS) is 
investigated using data taken in 1994 and 1995 
with the H1 detector at HERA at DESY, where 
820\,GeV protons collide with 27.5\,GeV positrons.
A measurement of the relative production rates of events 
with different jet multiplicities will lead to a better 
understanding of the underlying elementary processes
sensitive to perturbative QCD (pQCD) including its basic 
parameter, the strong coupling constant. 
\par
\begin{figure}[htb]
 \begin{center}
 \epsfig{file=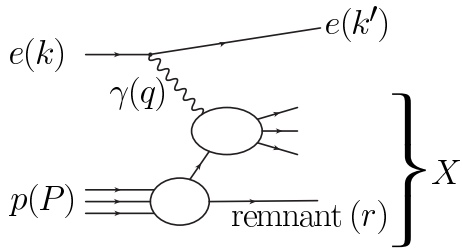,width=6.0cm}
 \caption[]{\em Sketch of neutral current (NC) deep inelastic 
           {\it ep} scattering (DIS) reaction $ep \rightarrow eX$, 
           where the four-momentum vectors are given in brackets.} 
 \label{fig1}
 \end{center}
 \end{figure}

The DIS reaction can be written as $e+p \rightarrow e+X$, 
where $X$ denotes an arbitrary hadronic final state. The 
process is sketched in figure~\ref{fig1}, in which the 
four-momentum vectors of the particles are given in brackets 
to define the kinematic quantities used later in the text.
The incident positron is scattered over a wide range of the 
negative squared four-momentum transfer~$\qq=-q^2=-(k-k^{\prime})^2$. 
The absorption of the virtual photon by the proton produces
a hadronic final state with a squared invariant mass of 
$W^2 = (q+P)^2$. 
\par
%
%
%
In the range of $\qq$ investigated in this paper, neutral current 
DIS reactions can be described as purely electro\-magnetic 
scattering of the positron off a quark via single photon 
exchange\footnote{In the $\qq$ range under study 
($\qq < 4000$~\gevq with the present statistics) 
effects due to the exchange of the intermediate vector 
boson $\zn$ can be neglected.}. To lowest order, this 
leads to (1+1) jet events. One jet is caused by the scattered 
quark, the other (+1) by the remaining part of the proton, 
the proton remnant, which can only be partially observed in 
the H1 detector. 
\par
\begin{figure}[htb]
 \begin{center}
 \epsfig{file=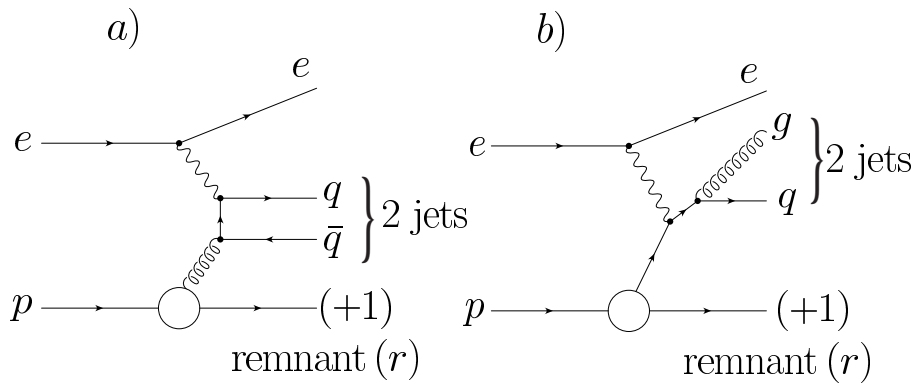,width=12.0cm}
 \caption[]{\em Generic diagrams of the QCD LO contributions to (2+1) jet 
            events in neutral current DIS.}
 \label{fig2}
 \end{center}
 \end{figure}

Higher jet multiplicities are described in the framework of 
perturbative QCD.
The corresponding cross sections are expressed as power series 
in the strong coupling constant $\al$, such that the leading 
order (LO) contribution to (2+1) jet events is of order $\oas$.
Two generic graphs contribute (figure~\ref{fig2}), where the 
hard scattering process is either gluon~(2a) or quark~(2b) induced. 
The resulting jets from the hard scattering are called ``current jets''. 
\par
%
%
%
By applying the modified JADE jet algorithm~\cite{jadcol,pseudo} to the 
data it has been shown in previous publications~\cite{H1alp,Zeus1},
that jet production at HERA is
dominated by the two diagrams of fig.~\ref{fig2} in a large 
region of phase space. 
This motivates a quantitative comparison to
pQCD predictions including real and virtual corrections
to the diagrams in question.
\par
Up to now the theoretical calculations~\cite{projet,disjet} available 
for such comparisons neglected terms containing the product of $W^2$ 
and the JADE jet resolution parameter $y_{cut}$ and thus were not valid 
for large values of $W^2$ accessed at low values of $\qq$. In recent NLO 
QCD-calculations~\cite{mepjet,disent} no such terms are neglected. Hence 
it is appropriate to compare more measurements with these new predictions
in a larger region of phase space, especially for low $\qq$ where higher 
statistics are available. Thus this analysis supersedes the previous H1 
publication~\cite{H1alp}.
\par
A necessary prerequisite to the analysis reported here is to define 
a phase space region in which NLO calculations are able to describe 
the data after correcting for hadronization and detector effects; the 
dependence of the (2+1) jet event rate on $\qq$, 
together with other hadronic control variables, should be well 
predicted by the fixed order calculation. 
\par
The presented quantitative pQCD analysis is based on
the previous H1 publication~\cite{H1alp}. 
The measured (2+1) jet event rates in different bins of $\qq$ are 
corrected for detector and hadronization effects to the parton level. 
Taking the known parton densities in the proton, 
values of $\al$ are extracted 
as a function of $\qq$ which 
fit best the corrected jet event rates. 
It is thus possible to investigate 
the scale dependence of the strong coupling constant
using only one observable, the (2+1) jet event rate, 
in a single experiment. 
Then $\almz$ is determined 
by performing a fit of the QCD parameter~$\Lambda$ 
in the renormalization group equation  
to the independent $\alq$ values. 
Here $M_Z$ is the mass of the neutral vector boson~$Z^0$. 
The use of a recent NLO QCD calculation program 
allows a study of the dependence of the obtained $\almz$ 
on the choice of the jet algorithm.
%

%
%
\section{H1 Apparatus}
\label{detector}
%
A detailed discussion of the H1 apparatus can be found
elsewhere~\cite{H1}. Here emphasis is put on describing the
main features of those detector relevant to this
analysis, which makes use mainly of the calorimeters and, to
a lesser  extent, of the central and backward tracking systems. 
The polar angle, $\theta$, is defined with respect to the
proton beam direction, which is identified with the 
$+z$-axis.

The inner part of the detector consists of the central tracking
chamber supplemented by a forward tracking detector and a
backward proportional chamber (BPC), covering the
polar angular ranges $ 25^\circ < \theta < 155^\circ$,
$ 7^\circ < \theta < 25^\circ$ and
$155^\circ < \theta < 175^\circ$, respectively.
The central and forward
tracking devices are used to determine the vertex position. 
The central tracking chamber and the BPC, together with the
vertex, are used to measure the positron scattering angle in the
central and backward region. The angular resolution achieved is 
better than 1~mrad. 
 
The energy of the scattered positron is measured by
the liquid argon~(LAr) calorimeter and in a
backward electromagnetic lead-scintillator calorimeter
(BEMC); the LAr calorimeter is also used for a measurement of the 
hadronic energy flow. 
A superconducting solenoid outside the LAr calorimeter provides 
a uniform magnetic field of $1.15$ T parallel to
the proton beam axis throughout the tracking region.
\par
The LAr calorimeter~\cite{LARC} extends over the polar angular range 
$4^\circ < \theta <  154^\circ$ with full azimuthal coverage. 
The scattered positron enters the LAr for events in which 
$\qq$ is larger than about 100 \gevq.
The calorimeter consists of an electromagnetic section with
lead absorbers, corresponding to a depth between 20 and 30 radiation
lengths, and a hadronic section with steel absorbers.
The total depth of the LAr~calorimeter varies between 4.5 and 8
hadronic interaction lengths. 
Test beam measurements of LAr~calorimeter modules have demonstrated
energy resolutions of $\isge\approx 0.12/\sqrt{E\comm{/ GeV}}\oplus 0.01$
for positrons~\cite{H1joerg}
and $\isge\approx 0.5/\sqrt{E\comm{/ GeV}}\oplus 0.02$
for charged pions~\cite{H1PI}.
The uncertainty of the absolute energy scale for positrons is 
$3 \%$ determined with studies based on the double angle 
method of kinematic reconstruction~\cite{Bent}. 
The hadronic energy scale has been verified from the balance 
of transverse momentum between the hadronic final state 
and the scattered positron in DIS events 
and is known to a precision of $4\%$~\cite{hadroenerscale}. 
\par
The BEMC~\cite{BEMC}, with a thickness of 21.7 radiation lengths, covers 
the backward region of the detector, $151^\circ < \theta < 176^\circ$.
It is mainly used to measure positrons from DIS processes at low $Q^2$. 
The acceptance region corresponds to  $Q^2$ values in the approximate 
range $5  \leq Q^2  \leq 100$ GeV$^2$. A resolution of 
$\isge\approx 0.10/\sqrt{E\comm{/ GeV}}\oplus 0.02$ has been achieved.
By adjusting the measured positron energy spectrum to the kinematic
peak the BEMC energy scale is known to an accuracy of $1\%$~\cite{BEMC}. 
Since 1995 the BEMC and BPC in the backward region have been replaced by 
a lead/scintillating fibre calorimeter (SPACAL)~\cite{spacal} and a 
drift chamber (BDC)~\cite{bdc}. 
%
%
%
%
\section{Event Selection}
\label{selection}
The data are divided into two sub-samples depending on the location 
of the detected positron in the BEMC or in the LAr calorimeter. 
The scattered positron is defined as either the most energetic cluster
detected in the BEMC or a contiguous cluster in the LAr whose energy 
in the electromagnetic section exceeds 80\% of the total cluster energy.
The used data samples correspond to integrated 
luminosities of 2.8~pb$^{-1}$ taken in 1994 for $\qq < 100$~\gevq 
(positrons in BEMC) and of 7.1~pb$^{-1}$ taken in 1994 and 1995 for 
$\qq > 100$~\gevq (positron in LAr calorimeter). All 
further selection criteria follow closely those used in the H1 
measurement of the proton structure function~$F_2$~\cite{f294}.

Events in the first sub-sample (BEMC) must satisfy
the following requirements:
\begin{enumerate}
\item The four-momentum transfer $Q^2$ must be between 10 and  100 GeV$^2$.
\item The energy  $E_e^{'}$ of the scattered positron
      must be greater than  14 GeV, corresponding to 
      a fractional positron energy loss in the proton rest-frame
      of $y \, = \, qP/kP \, \lsim \, 0.5 $ (see fig.~\ref{fig1} for 
      the meaning of four-momentum vectors), 
      thus eliminating background from photo-production,
      where a hadron is misidentified as a positron, 
      to a negligible level ($<$1\%). 
      In addition this cut suppresses radiative DIS events.
\item The angle of the scattered positron must be in the range
      $160^\circ  <\theta_e< 173^\circ $ 
      to ensure that it is well contained in the BEMC. 
\end{enumerate}

Events in the second sub-sample (LAr) must satisfy the
following requirements:
\begin{enumerate}
\item $\qq$ must be greater than 100~GeV$^2$. 
\item The energy of the scattered positron~$E_e^{'}$ 
      must be greater than 11~GeV, 
      $y$ less than 0.7, and
      the quantity $\delta=\sum_{clusters}(E - P_z)$ 
      (with $E$ and $P_z$ being  the energy and longitudinal 
      momentum of the calorimetric energy clusters) is 
      required to be within $38$ GeV $ < \delta < 70 $ GeV. 
      These cuts exclude efficiently photo-production events ($<$1\%), 
      in which the scattered positron is lost down the beam pipe. 
      As for the cut in $y$ discussed above, the lower $\delta$-cut 
      further reduces radiative DIS events.
\item The angle $\theta_e$ should fulfill 
      $10^\circ < \theta_e< 150^\circ $ to ensure
      that the positron is contained in the
      LAr~calorimeter and to avoid the transition region between the LAr
      and the BEMC.
\end{enumerate}

Two additional requirements were imposed on both sub-samples:
\begin{enumerate}
\item  An event vertex from charged tracks 
       within $\pm 30$ cm around the maximum, $z_{mean}$,  
       of the vertex distribution is required; $z_{mean}$ was located at 
       $z=+5$~cm (+1~cm) in 1994 (1995). 
\item  $W^2 > 5000$~GeV$^2$, where $W^2$ is calculated using the 
       double angle method. This method relies on the angle
       of the scattered positron and an inclusive angle of the 
       hadronic final state, and is independent of the jet classification. 
       The $W^2$ cut ensures enough hadronic activity in the detector 
       and is equivalent to a cut on the lepton variable $y$ $(y>0.055)$. 
\end{enumerate}

In both samples $Q^2$ and $y$ are calculated
from the energy and angle of the scattered positron.
The selected event sample contains 112985 events with $Q^2<100$ GeV$^2$ and 
5410 (8280) events with $100$~GeV$^2 < Q^2 < 4000$~GeV$^2$ selected 
from the data taken in 1994 (1995). 
%

%
%
%
%
\section{Jet Reconstruction}
\label{jetreco}
%
All events are subject to a jet classification 
using the energy clusters in the LAr calorimeter always
excluding the scattered positron.
In order to compare the measurement to QCD predictions  
the same jet classification is applied to 
Monte-Carlo events after passing the full H1 detector 
simulation and after using the same reconstruction 
algorithm as for the data. 
The influence of hadronization and detector effects 
on the number of reconstructed jets at this ``detector 
level'' is studied by determining in addition the 
jets at the parton and the hadron level. 
At the ``parton level'' the coloured partons 
generated according to pQCD calculations 
will form the jets, whereas at the ``hadron level'',
the fully generated hadronic final state 
is analysed. 

The analysis presented here is based on a modified JADE jet 
algorithm~\cite{jadcol}. In order to take into account the invisible 
part of the proton remnant~$r$ the modified JADE jet algorithm includes 
a pseudo-particle in the clustering procedure representing the missing 
longitudinal momentum in the event~\cite{pseudo}. The JADE algorithm is 
applied in the laboratory frame. Two objects $i,j$ with a squared mass 
$\mij$ are combined to one new object if $y_{ij} = \mij/\mref^2 < y_{cut}$, 
where $y_{cut}$ is the resolution parameter. The scale~$\mref$ is taken 
to be the invariant mass~$W$ of the hadronic system $X$, where $W$ is 
calculated as the invariant mass of the four-vector sum of all hadronic
objects. From studies of the resolution parameter~\cite{richard,h1mujra94}, 
$y_{cut}$ is set to 0.02. This choice ensures satisfactory correlations 
between jets reconstructed at the parton and the detector level, and high 
statistics of (2+1) jet events, keeping the contribution of (3+1) jet 
events below 4\%. 

Besides the modified JADE jet algorithm two of its variants, the E0 and
the P algorithm, are investigated, which differ in the combination of two 
objects $i$ and $j$ to a new object~$k$. 
For the JADE algorithm the new four-momentum 
vector~$p_k$ is given by $p_k = p_i + p_j$, whereas for the E0 and P 
algorithm the new four-momenta are massless and given by
$p_k = [E_i+E_j,
       (E_i+E_j)\cdot\frac{\vec{p}_i+\vec{p}_j}{|\vec{p}_i+\vec{p}_j|}]$
and
$p_k = [|\vec{p}_i+\vec{p}_j|,\vec{p}_i+\vec{p}_j]$, respectively.
The resolution criterion for all of these algorithms is given by 
$y_{ij} = 2 E_i E_j (1-\cos\theta_{ij})/W^2$, where $E_i,E_j$ 
denote the energies and $\theta_{ij}$ is the angle between the two 
objects~$i,j$ under consideration. 
As long as there is no restriction on the momenta of 
the resulting jets, the JADE and E0 algorithm are equivalent in 
the theoretical NLO (2+1) jet cross section calculations. 
For these two algorithms, only the (1+1) jet cross section is 
affected differently by cuts on the current jet angles. 
The difference, which is marginal, is caused by events 
where the clustering of all three partons and the proton 
remnant leads to a (1+1) jet event. 

%
%
\section{Outline of the QCD Analysis}
\label{method}
%
In order to study the applicability of pQCD quantitatively, the (2+1) 
jet event rate for different bins of $\qq$ (see table~\ref{tab:rate}) 
is chosen~\cite{H1alp,richard} to determine 
$\almu$ as a function of the renormalization scale\footnote{The 
factorization scale~$\mu_f^2$ is also set to $\qq$.}~$\murq$,
which is taken to be $\qq$.
The NLO QCD calculation can be used to calculate an expression for
$R_{2+1}(\qq,y_{cut})$ in terms of a power series in $\al(\qq)$ with
coefficients $a(\qq,y_{cut})$ and $b(\qq,y_{cut})$:
\[
R_{2+1}(\qq,y_{cut}) = \frac{\sigma_{2+1}(\qq,y_{cut})}{\sigma_{norm}(\qq,y_{cut})} 
                     = a(\qq,y_{cut}) \al(\qq) + b(\qq,y_{cut}) \al^2(\qq)
                       + {\cal O}(\al^3).
\]
Taking the variation of $\alpha_s$ with $\qq$ to be negligible over each
bin of $\qq$, to $\oasd$ this equation reduces to a
quadratic form for $\alpha_s$ with coefficients determined from the
integrals of $a(\qq,y_{cut})$ and $b(\qq,y_{cut})$ over each $\qq$ bin. This
quadratic expression can then be solved for $\alpha_s$ using the measured
jet event rates for different $\qq$.

The cross section used for normalization, $\sigma_{norm}(\qq)$,
is taken as the sum of jet cross sections, where one or two current jets 
($\sigma_{1+1}$ for (1+1) and $\sigma_{2+1}$ for (2+1) events) 
are found in the acceptance region. Experimentally $R_{2+1}$ is 
obtained as $N_{2+1}/(N_{2+1} + N_{1+1})$, where $N_{i+1}$ denotes the 
number of events with $i$ observed current jets. 

The theoretical $\oasd$ Monte Carlo integration programs,
MEPJET~\cite{mepjet} and DISENT~\cite{disent}, used for the prediction 
of jet cross sections are only available at the parton level 
and not as event generators. Both programs give the same results to 
within 1\% precision. Consequently the relation between 
observed jets in the detector and the underlying partonic jet 
configuration has to be obtained in form of correction factors from 
a LO event generator which includes a simulation of higher-order
effects and hadronization.

The MEPJET Monte Carlo program is used to compare the NLO prediction 
with the measured jet event rates and distributions of kinematic 
variables. It is necessary to apply to the NLO Monte Carlo the same
phase space restrictions in terms of positron variables and 
jet quantities as is done in the experimental event selection.
The forward direction is the most crucial region for the comparison. 
The emission of multi-gluon radiation from initial state partons 
entering the hard scattering process 
causes a higher reconstructed jet multiplicity. The influence of 
multi-gluon radiation increases with decreasing $\qq$
because at low $\qq$ very low values of 
the scaling variable $x_B=\qq/2qP$ are reached, which 
increases the permitted phase space ($\xi > x_B$) 
for the momentum fraction~$\xi$ of the proton carried by the 
parton initiating the multi-gluon radiation. 
This initial state radiation is naturally not taken into account 
by a fixed order theory (as in MEPJET). 
As a consequence the jet event rate which is predicted by the NLO
calculations is much smaller than the measured event rate 
without further cuts, particularly at low values of $Q^2$. 
Therefore the event generator Monte Carlo is not only used to 
calculate the correction factors but also to determine cuts 
necessary to suppress the influence of multi-gluon radiation 
and hadronization effects. 

To define such a phase space region where the NLO QCD prediction can 
describe the partonic scattering underlying the data events, an event 
generator is necessary, which has a parton level comparable to the partonic 
configuration of the NLO QCD calculation and describes the data.
Therefore the acceptance region for the current jets is chosen  
using the following criteria: 
\begin{itemize}
  \item The dependence of the jet event rate on $\qq$ predicted by the 
        NLO calculation should be in good agreement with 
        the prediction of the event generator at the parton level. 
        For a reasonable range of the QCD parameter 
        $\Lambda^{(4)}_{\scriptstyle\overline{\it{MS}}}$ 
        as used in MEPJET\footnote{In MEPJET the effective number
        of flavors~$n_f$ contributing to the QCD parameter 
        $\Lambda^{(n_f)}_{\scriptstyle\overline{\it{MS}}}$ 
        (the calculation assumes zero quark masses)
        is fixed to 5; but in the text presented here always the corresponding 
        $\Lambda^{(4)}_{\scriptstyle\overline{\it{MS}}}$ 
        will be given.} 
        ($150 < \Lambda^{(4)}_{\scriptstyle\overline{\it{MS}}} < 400$~MeV)  
        also the absolute values of the jet event rate should be close to the 
        predicted values from the event generator 
        in each bin of $\qq$.
        In addition other reconstructed jet variable distributions at
        partonic level, such as jet angles, should be in acceptable
        agreement with the NLO calculation.
  \item The measured (2+1) jet event rates and the other 
        data distributions of the chosen jet variables 
        should be described 
        by the event generator model at the detector level. 
\end{itemize}
It is assumed that if these requirements are fulfilled the correction
factors from measured jets to the NLO parton configuration can be provided
by the event generator.

As event generator the program LEPTO\,6.5~\cite{LEPTO65} is used.
LEPTO enables event generation by matching leading log parton showers 
to the LO matrix elements (MEPS). In the event generator 
a momentum scale~$Q_0$ specifies the termination of the parton shower, 
that is the boundary between the perturbative and non-perturbative stages 
of jet development. For this analysis, this scale is taken to be 
$1$~\gev~as motivated in~\cite{berger}. 
It has been demonstrated that this combination of matrix elements and
parton showers gives a good description of the data in a
wide range of the phase space 
specified by the cuts below~\cite{H1alp,h1mujra94,christian,fabrice}. 
Nevertheless, there is some arbitrariness in the definition of the parton 
level. Investigations have shown that the determination of $\alpha_s$ can
depend on the definition of the parton level at the order of a few per 
cent~\cite{opal1990,mweber}. The interpretation of these investigations 
is, however, difficult. It can be argued that this sensitivity is due to 
effects of missing higher order terms which may well be included in the 
systematic errors due to the uncertainties of the available QCD models, in 
the renormalisation scale and in the choice of the used jet algorithm 
(see below). Accordingly, no additional source of systematic error is 
included for this arbitrariness.

For the selected phase space region, in each $\qq$ bin a correction factor 
to the experimental jet event rate is calculated using the LEPTO\,6.5 
generator from the ratio of the jet event rates obtained at the parton level 
after parton showering and at the detector level. 
In the same way also correction factors from the detector to the 
hadron level are determined.
A possible model dependence introduced by this prescription is 
investigated by using the event generators 
HERWIG~\cite{HERWIG} and ARIADNE~\cite{ariadne} 
and its effect is included in the systematic error. 

The theoretical predictions are calculated using parton density 
parameterizations with the factorization scale~$\mu_f^2$ given by $\qq$. 
The parton densities are usually obtained via evolution from a low 
lying scale using a value $\almz$ as input parameter, which may 
differ from that extracted in the analysis presented here. It will 
be shown that the error introduced by this inconsistency is small 
in the kinematic region considered. 
%

%
%
%
%
\section{Phase Space Selection}
\label{jade}
%
The events selected according to section 3 are subject to a jet 
classification applying the
modified JADE jet algorithm as discussed in section~\ref{jetreco}. 
A resolution parameter of $y_{cut}=0.02$, in combination with the cut 
$W^2 > 5000$ GeV$^2$ (section~\ref{selection}), implies an invariant 
mass squared of the two current jets which is greater than 100~GeV$^2$.
After the jet classification a cut in the maximum jet angle 
($\theta_{max}=145^\circ$) is applied to ensure that the jets are 
measured in the LAr calorimeter. 

A cut at moderate values of $\qq$ ($\qq > 40$ \gevq) ensures that 
$R_{2+1}$ is well described by LEPTO. In addition it avoids phase 
space regions where the variation of the QCD parameter~$\Lambda$ is 
more important in the parton density functions than in the 
hard-scattering cross sections~\cite{chyla}. Furthermore, since this cut 
does not suppress parton showers sufficiently, additional requirements are 
necessary. Parton showers produce jets 
with predominantly small angles ($\tj$) relative to the incident 
proton direction and small values of 
%
 \begin{figure}[b] \unitlength 1mm 
 \begin{picture}(200,95) \put(64.5,92){(\,a\,)} \put(149.5,92){(\,b\,)}
 \put(-4.5,0){\epsfig{figure=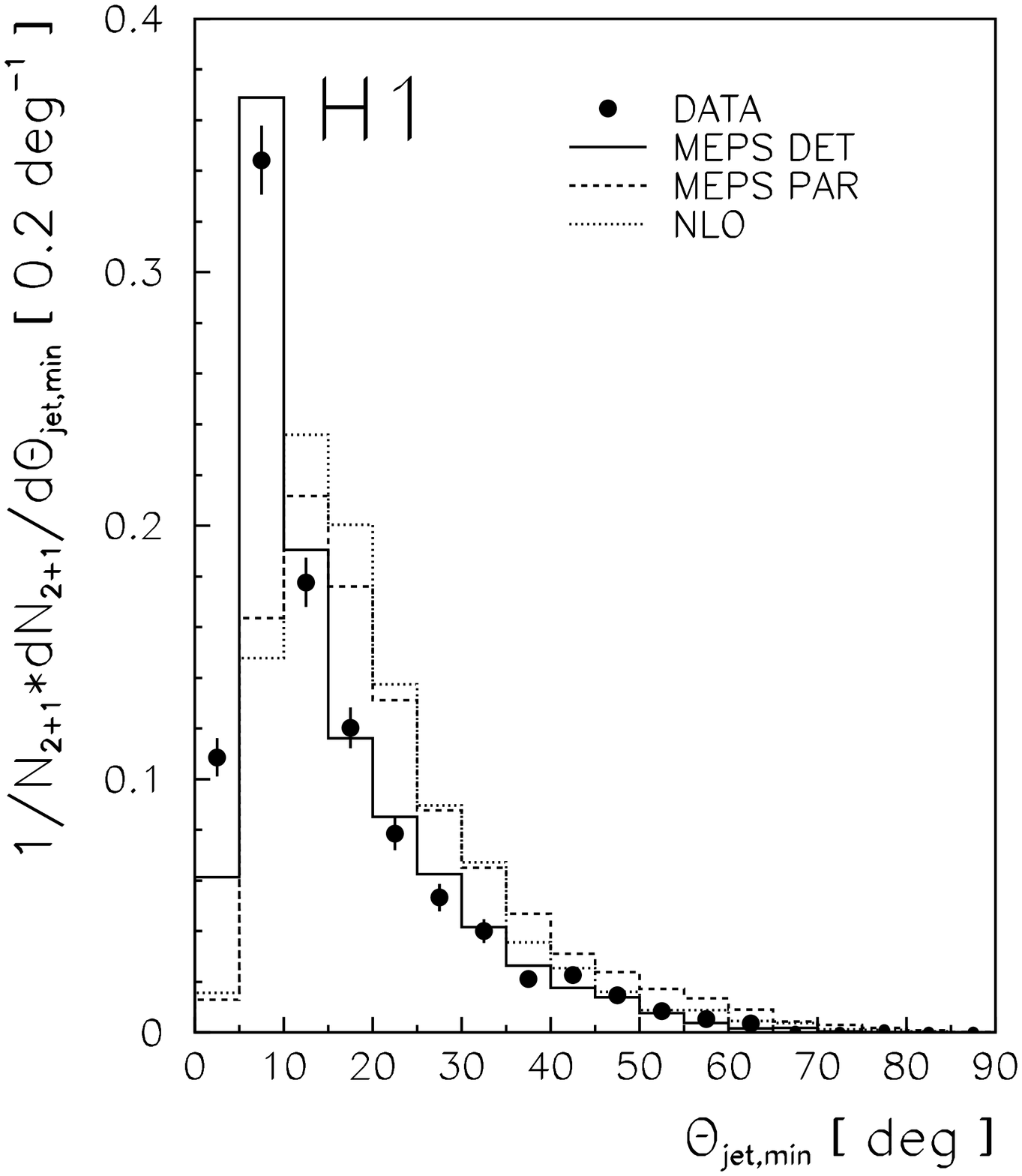,width=80mm,angle=0}}
 \put(80.5,0){\epsfig{figure=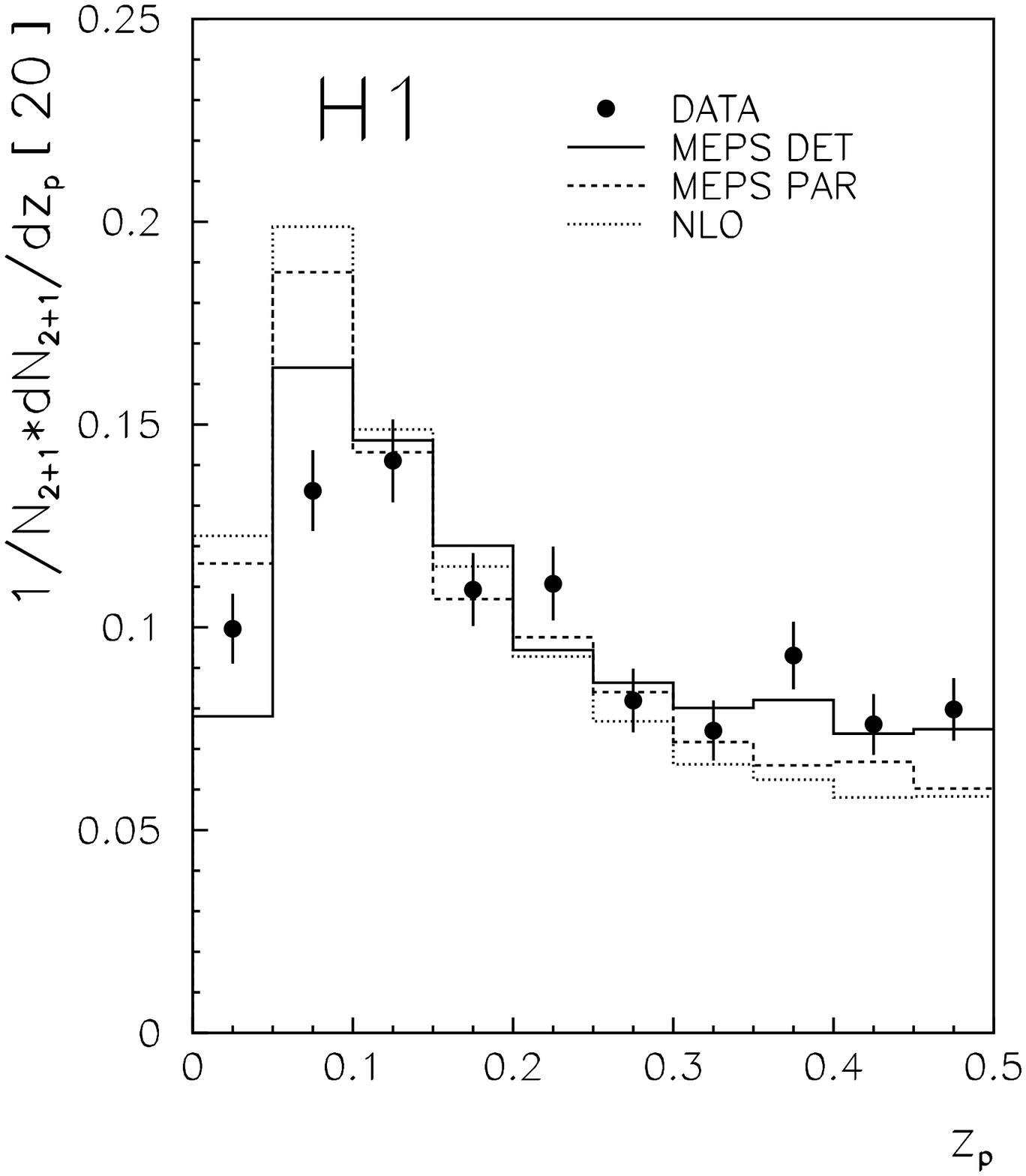,width=80mm,angle=0}}
 \end{picture} \normalsize \unboldmath
 \vspace{-2.cm}

 \caption[dum]{\it
a) The $\theta_{jet,min}$ distribution of the (2+1) jet events for 
Q$^{\mb{2}} >$~40~GeV$^{\mb{\,2}}$, $z_p >$~0.1 and 
$\theta_{max}=$~145$^{\,\circ}$; 
b) The $z_p$ distribution of the (2+1) jet events for 
Q$^{\mb{2}} >$~40~GeV$^{\mb{\,2}}$, 10$^{\,\circ} 
< \tj <$~145$^{\,\circ}$. 
The JADE algorithm is used to define the jets; in both figures 
the data are represented by the points with statistical error bars only. 
The solid line shows the prediction of the MEPS model at the 
detector level and the dashed line the prediction of the same
model at the parton level. 
The NLO calculation 
($\Lambda^{(\mb{4})}_{\scriptstyle\overline{\it{MS}}}=$~230~MeV) 
is given by the dotted line. 
All curves are normalized to the number of (2+1) jet events.}
\label{zp}\end{figure}
\[
 z_p = \min_{i=1,2} \frac{ E_{jet,i} (1 - \cos\theta_{jet,i}) }
                    { \sum_{j=1}^{2} E_{jet,j} (1 - \cos\theta_{jet,j}) } \:.
\]
The values of $z_p$ determined  
by the measured energies~$E_{jet,j}$ and angles~$\theta_{jet,j}$ 
$(j=1,2)$ of the jets in the laboratory frame 
are in the range $0<z_p<0.5$. 
The $\tj$ and $z_p$ distributions are strongly correlated but are projections 
of different regions of the jet phase space in the $(\tj, z_p)$-plane. 
Both are used in this analysis to restrict the jet acceptance region. 

In fig.~\ref{zp}a the distribution of the smallest jet angle 
($\theta_{jet,min}$) relative to the incident proton direction
in a ($2+1$) jet event using the JADE algorithm 
is plotted for $\qq>40$ \gevq \, and $z_p>0.1$.
The curves are normalized to the number of events. 
It can be seen that above $\theta_{jet,min}=10^\circ$ 
the data are well reproduced by the 
MEPS model after detector simulation.
In addition the MEPS model at the parton level and 
the MEPJET calculation (with MRSH as parton density parameterization 
and $\lav =$ 230 MeV) 
agree reasonably well with each other. 
Both the MEPS model and the NLO calculation 
predict however fewer ($2+1$) jet events with 
smallest $\theta_{jet,min}$ values 
than measured in the data. 
A $\tj$ cut suppresses parton showers 
and hadronization effects efficiently. 
The $z_p$ distribution is shown 
for $\qq>40$ \gevq and $\tj> 10^\circ$ in fig.~\ref{zp}b. 
The final phase space selection fulfilling the criteria described in
section~\ref{method} is therefore given by the requirements
$\qq>40$ \gevq, $\tj > 10^\circ$ and $z_p >0.1$, and is hereafter
referred to as the central cut scenario. With these requirements the
final event sample contains 1038 (2+1) jet events of which 689 have
$\qq>100$ \gevq.

Due to effects of the finite resolution of energy and angles
in the calculation of $m_{ij}^2$, there are migrations between jet 
classes at the levels of the detector, hadrons and partons.
The efficiency and purity of the selected BEMC (LAr) sample 
are 27\% (40\%) and 57\% (69\%) with respect to the 
jet classification at the parton level and the above cuts~\cite{christian}. 
%
 \begin{figure}[t] \unitlength 1mm 
 \begin{picture}(200,95) \put(64.5,92){(\,a\,)} \put(149.5,92){(\,b\,)}
 \put(-4.5,0){\epsfig{figure=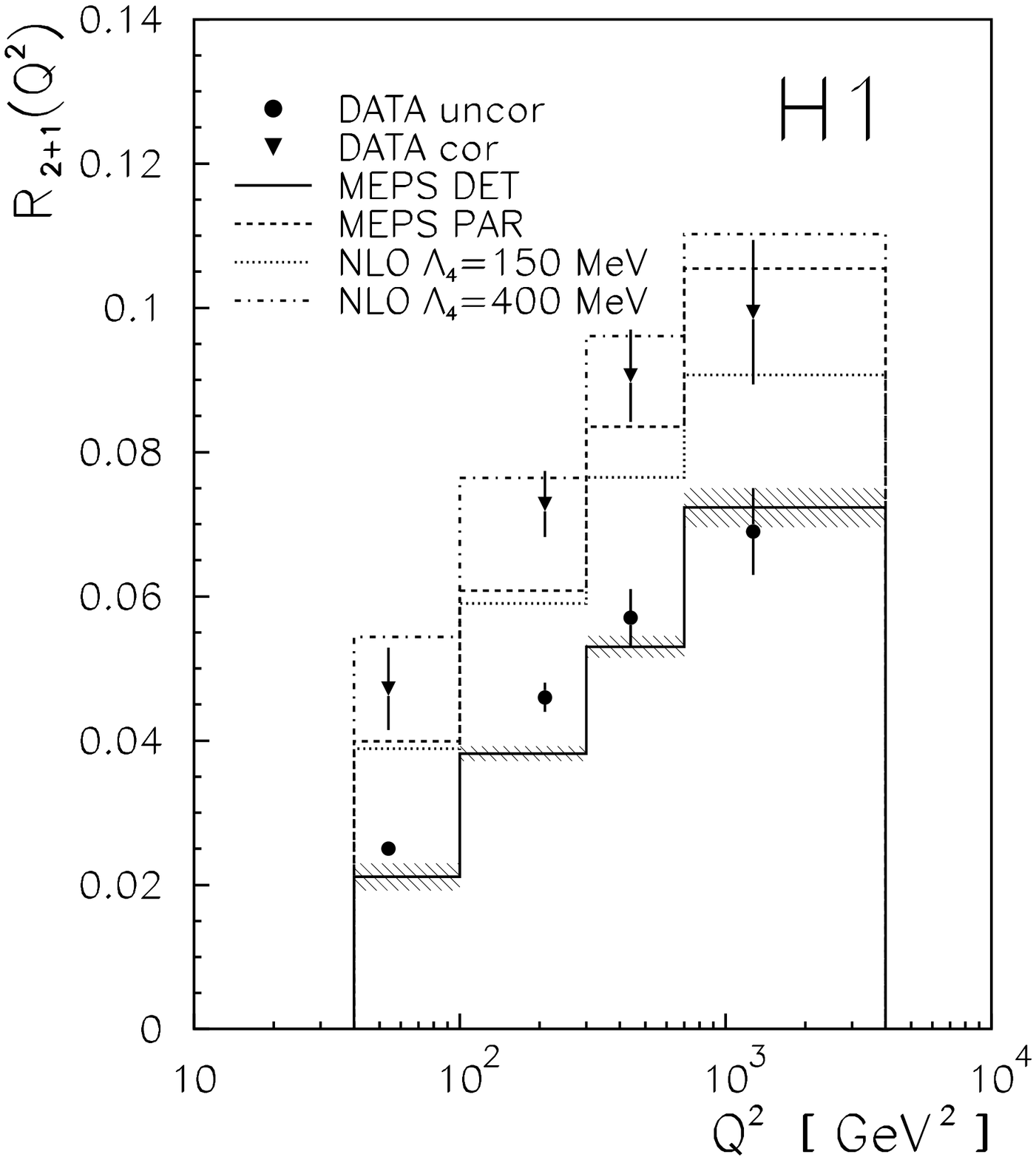,width=80mm,angle=0}}
 \put(80.5,0){\epsfig{figure=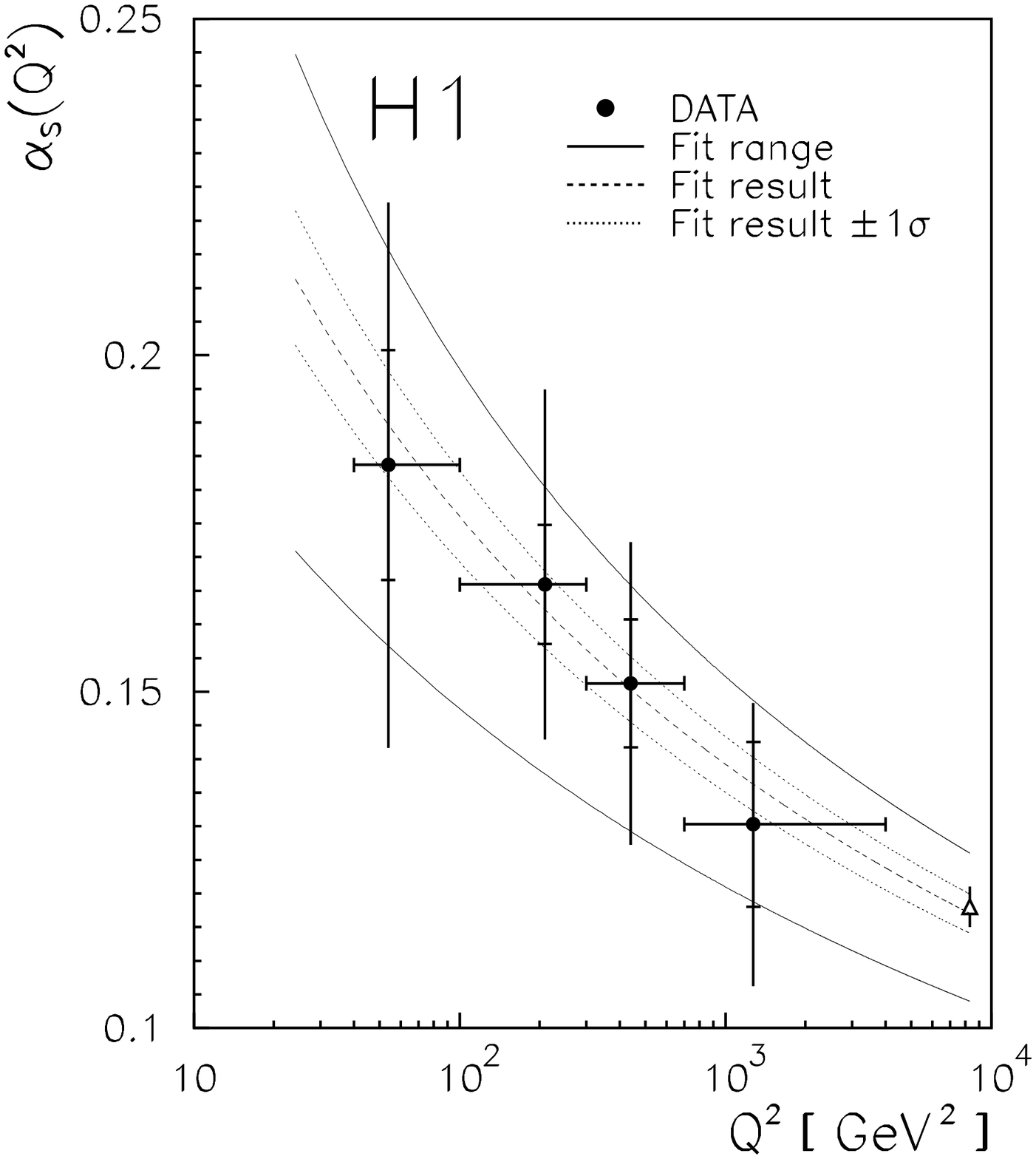,width=80mm,angle=0}}
 \end{picture} \normalsize \unboldmath
 \vspace{-2.cm}

 \caption[dum1]{\it The $\al$ determination 
                at four values of Q$^{\mb{2}}$ and the evolution to 
                the scale Q$^{\mb{2}}=$ M$_{Z}^{\mb{\,2}}$ 
                using the modified JADE jet algorithm.
                The data points are placed at the average Q$^{\mb{2}}$ in 
                each bin. \newline
 a) The measured $R_{\mb{2+1}}$(Q$^{\mb{2}}$) (points) and the jet event 
    rates corrected to the parton level (triangles) together with the 
    prediction of the MEPS Monte Carlo at the detector level including
    statistical error bands (solid line), parton level (dashed line) and 
    the MEPJET NLO calculations (dotted and dashed dotted lines) for two 
    values of $\Lambda^{(\mb{4})}_{\scriptstyle\overline{\it{MS}}}$.
    The event rates are determined for 
    $y_{cut}=$~0.02, Q$^{\mb{2}} >$~40 GeV$^{\mb{\,2}}$, 
    $\theta_{max}=$~145$^{\,\circ}$,
    $\tj>$~10$^{\,\circ}$ and $z_p>$~0.1. The predictions are based
    on the MRSH parton density parameterization. \newline
 b) The $\al$ values derived from the jet event rates as function of 
    Q$^{\mb{2}}$ with statistical errors from the data and the correction 
    factors as inner error bars, and total errors as full error bars.
    The width of the Q$^{\mb{2}}$ bins are indicated by the horizontal bars. 
    The fit result (dashed line) and the 1 s.d.\ errors (dotted lines) 
    are based on the individual $\al$(Q$^{\mb{2}}$) values with 
    their statistical errors only. The solid lines represent the 
    obtained range for $\alpha_s$(M$_{Z}^{\mb{\,2}}$) taking into 
    account the systematic uncertainties, except the jet algorithm error. 
    The open triangle at 
    Q$^{\mb{2}}=$ M$_{Z}^{\mb{\,2}}$  indicates the world average 
    $\alpha_s$(M$_{Z}^{\mb{\,2}}) =$ 0.118 $\pm$ 0.003~\cite{PDG96}.}
\label{al3}\end{figure}

The influence of radiative QED corrections 
on the (2+1) jet event rate is studied with the event 
generator DJANGO~\cite{django} and is negligible. 
In this acceptance region the NLO corrections
to the LO calculation are less than $30\%$  
in the investigated range of $\qq$ and the Bjorken
variable $x_B$.  

Using the JADE algorithm the measured (2+1) jet event rate $R_{2+1}$ 
for $\qq>40$ \gevq, $\tj>10^\circ$, $z_p>0.1$ is shown
in fig.~\ref{al3}a. 
The data above $\qq=40$ \gevq are divided into four bins
containing approximatively equal numbers of (2+1) jet events.
The measurements compare reasonably with 
the MEPS Monte Carlo at the detector level.  
In fig.~\ref{al3}a it can also be seen that the jet event rate at the parton 
level in the MEPS model lies well between the MEPJET prediction 
for $\lav=150$ MeV and $\lav=400$ MeV.
%

%
%
%
%
\newpage
\section{Determination of the Strong Coupling Constant}
\label{qqcdt}
%
%
\begin{table}[t]
\begin{center}
{\small
\begin{tabular}{|c|c|c|c|c|c|}\hline 
\va 
$\qq$ [\gevq] & 
$R_{2+1}$ [$10^{-3}$]& 
$R_{2+1}^{had}$ [$10^{-3}$] & 
$R_{2+1}^{par}$ [$10^{-3}$] & 
$f(det \rightarrow had)$ &  
$f(det\rightarrow par)$ \\\hline\hline
\va  
40-100 &   
$25\pm1 ^{+\,2}_{-\,2}$ & 
$33\pm4 ^{+\,4}_{-\,3}$ & 
$47\pm6 ^{+\:\:4}_{-\:\:9}$ & 
$1.32\pm0.15+0.14$ & 
$1.88\pm0.21-0.03$ \\\hline
\va 
100-300 &  
$46\pm2 ^{+\,3}_{-\,4}$ & 
$60\pm4 ^{+\,4}_{-\,5}$ & 
$73\pm5 ^{+\,\:\:9}_{-\:\:8}$ & 
$1.30\pm0.04+0.00$ & 
$1.59\pm0.05-0.12$ \\\hline
\va 
300-700 &  
$57\pm4 ^{+\,3}_{-\,0}$ & 
$72\pm5 ^{+\,4}_{-\,1}$ & 
$91\pm6 ^{+\,\:\:5}_{-\,13}$ & 
$1.26\pm0.04-0.02$ & 
$1.60\pm0.05-0.14$ \\\hline
\va 
700-4000 & 
$69\pm6 ^{+\,4}_{-\,1}$ & 
$82\pm8 ^{+\,5}_{-\,2}$ & 
$99\pm10 ^{+\:\:8}_{-11}$ & 
$1.19\pm0.06-0.02$ & 
$1.43\pm0.07-0.09$ \\\hline
\end{tabular}
}
\caption{\it The measured (2+1) jet event rates $(R_{\mb{2+1}})$ 
         with statistical and systematic errors in 
         different bins of $\qq$ for the central cut scenario 
         using the modified JADE algorithm and corrected to the hadron 
         $(R_{\mb{2+1}}^{had})$ and parton level $(R_{\mb{2+1}}^{par})$. 
         In addition the correction factors from the detector 
         to the hadron $(f(det \rightarrow had))$ and 
         parton $(f(det\rightarrow par))$ level are included  
         with errors due to Monte Carlo statistics
         and the model uncertainty estimated by ARIADNE. 
         The statistical errors on the measured $R_{\mb{2+1}}$ are given by 
         the data only, whereas the errors on the corrected (2+1) 
         jet event rates $(R_{\mb{2+1}}^{had},R_{\mb{2+1}}^{par})$ include 
         also the statistical errors from the Monte Carlo. 
         The systematic errors on $R_{\mb{2+1}}$ correspond to the 
         uncertainty on the hadronic energy scale. 
         The systematic errors on $R_{\mb{2+1}}^{had}$ take into account 
         the hadronic energy scale and model uncertainty. 
         For $R_{\mb{2+1}}^{par}$
         also the effects of parton showers, hadronization parameters 
         and QCD Monte Carlo models were taken into account 
         similar to the error estimation on $\al$ as described later 
         in the text.} 
\label{tab:rate}
\end{center}
\end{table}

The correction factors $f(det\rightarrow par)$, necessary to obtain from the 
measured jet event rate~$R_{2+1}$ the rate at the parton level 
$(R_{2+1}^{par})$, are given in table~\ref{tab:rate} for the different 
$\qq$ bins shown in fig.~\ref{al3}a. They are between 1.4 and 1.9.
Also the factors $f(det\rightarrow had)$ for 
correcting the measured event rates to the hadron level are 
included in table~\ref{tab:rate}. 
It can be seen that hadronization and detector effects lead 
to lower observable (2+1) jet event rates than at the parton level 
due to resolution effects. 
The statistical errors on the corrected event 
rates in table~\ref{tab:rate} 
include the statistical error from the data as well as from the 
calculation of the correction factors using the Monte Carlo generator. 
For all calculations the parton density distributions were parameterized
using MRSH~\cite{MRSH}.

The experimental event rates corrected to the parton level $R_{2+1}^{par}$ 
are also plotted in fig.~\ref{al3}a for the standard set 
of cut parameters (see above) using the JADE algorithm. 
They are converted into four values of $\alq$.
The QCD parameter~$\Lambda^{(4)}_{\scriptstyle\overline{\it{MS}}}$ 
in the 2-loop solution of the renormalization 
group equation is fitted~\cite{Minuit} to these 
extracted values of $\alq$ 
taking into account their statistical errors only. 
The results, together with the $\pm 1$ s.d.\ parameter 
error of the fit, are shown in figure~\ref{al3}b.  
The observed $\qq$ dependence is compatible with the expected
theoretical evolution. The numerical result at $\qq = \mzq$ is 
$\almz = \alcena$. 

To obtain the systematic uncertainty on $\almz$ the fit 
is repeated for various other scenarios changing cuts 
and parameters relevant for the jet analysis. 
This procedure automatically takes into account 
the correlation between the four bins in $\qq$. 
All systematic errors discussed in detail below 
are summarized in table~\ref{errors}.

%
\begin{table}[b]
\begin{center}
\begin{tabular}{|l|c|c|}\hline
 Source                     & Variations & $\Delta \alpha_{s}(M^{2}_Z)$
 \\\hline\hline
 Statistics                 &&$ \pm\, 0.003 $ \\\hline\hline  
 $\tj$ and $z_p$ cut        & variation in $\tj$ and $z_p$, see text 
                            & $ ^{+\,0.004}_{-\,0.004}$ \\\hline  
 Hadronic energy scale      &$\pm 4\%$ &$ ^{+\,0.005}_{-\,0.003} $ \\\hline\hline
                            & LEPTO 6.5& $ ^{+\,0.004}_{-\,0.002}$\\\cline{2-3}
 Model dependence           & HERWIG 5.9 & $ -\, 0.007  $\\\cline{2-3}
                            & ARIADNE~4.08 & $ -\, 0.005  $\\\hline\hline
 Parton density functions,  &
 MRSAp-201, GRV92, CTEQ-4A1      & \\  
 $\Lambda^{(4)}_{\scriptstyle\overline{\it{MS}}}$ 
 used in the PDF's & 0.15\,,\,0.40 GeV& \rb{$^{+\,0.003}_{-\,0.002}$} \\\hline
 Renormalization and factorization scale &$\mu^2_i=1/4,4\,\qq$
 i=f,r &$ ^{+\,0.003}_{-\,0.007}$\\\hline
 $y_{cut}$                  & 0.015, 0.025 & $ ^{+\,0.002}_{-\,0.003} $ \\\hline\hline
 jet algorithms             & E0         & $ +\, 0.000  $\\         
                            & P          & $ +\, 0.006  $\\\hline
 \end{tabular}
 \caption[]{\it The determination of various systematic uncertainties. 
            The central cut scenario and the MRSH parton density 
            parameterization are used as reference.}
 \label{errors}
\end{center}
\end{table}
%

In order to determine the systematic experimental error the
following kinematic cuts are investigated:
\begin{itemize}
  \item The cut in $z_p$ is varied between 0.05 and 0.15 for
        $\tj > 5^\circ,10^\circ,15^\circ$ leaving the other parameters 
        at their central values. 
        In addition $\theta_{max}$ is varied between
        $140^\circ$ and $150^\circ$ leaving the other
        parameters at their central values. 
        The resulting error on $\almz$ as determined from the 
        maximum spread of all fit results
        is $+0.004$ and $-0.004$. 
  \item The hadronic energy scale of the LAr calorimeter 
        in the data reconstruction 
        is varied by $\pm 4\%$. The resulting error on $\almz$ 
        is $+0.005$ and $-0.003$. 
\end{itemize}
The systematic experimental error ($+0.006$ and $-0.005$) on $\almz$ 
is taken as the quadratic sum of the two errors given above.
Effects due to the uncertainty of the energy of the scattered positron 
and the variation of kinematic cuts defined in section~3 
are negligible. 

A further source of systematic uncertainty stems from the dependence
of the correction factors on the Monte Carlo models used to calculate them.
These uncertainties are analyzed employing the following procedure:
\begin{itemize}
 \item The LEPTO Monte Carlo uses the 
       JETSET~\cite{jetset} fragmentation model. 
       Several sets of parameters describing the hadronization 
       and fragmentation as fitted by the LEP 
       experi\-ments~\cite{lepexp} are used to 
       study the stability of the correction factors 
       from the hadron to the parton level. 
       The widths of the Gaussian distributions for the 
       transverse momenta of partons in the proton 
       ($k_T$ (\,PARL(14)\,) and $\sigma_{pt}$ (\,PARL(3)\,))
       are varied between $350 - 700$~MeV and $440 - 700$~MeV, 
       respectively;  
       the fragmentation parameter $a$ of the LUND 
       fragmentation functions is varied 
       between 0.1 and 1; 
       the width~$\sigma$ (\,PARJ(21)\,) in the gaussian $p_x$ 
       and $p_y$ transverse momentum distributions for primary 
       hadrons is changed from the default value (360~MeV) 
       to 250 and 450~MeV. 
       In addition the momentum scale which determines the termination of
       the parton shower ($Q_0$-parameters PYPA(22) and PARJ(82) in LEPTO
       for the initial and the final state partons respectively) are
       varied between 1 and 2~GeV and between 1 and 4~GeV, respectively.
       Finally the lowest value of the invariant mass squared $\hat{s}$ of 
       the two generated partons, for which the LO ME calculations are 
       carried out, is changed from 4~\gevq to 25 \gevq. The resulting 
       fitted $\almz$ values are in the range 0.115 to 0.121.
 \item Instead of LEPTO, the HERWIG~\cite{HERWIG} Monte Carlo model with
       its different treatment of parton showering and 
       hadronization is applied to calculate the (2+1) jet event rate on 
       parton and hadron level. 
       HERWIG is able to describe the shape of the measured $R_{2+1}(\qq)$  
       distribution. 
       Using the same corrections for detector effects as determined with
       the LEPTO generator, the fit leads to a reduction of $\almz$ by
       0.007. 
 \item The event generator ARIADNE~4.08~\cite{ariadne} is also
       used to study the model dependence. This Monte Carlo model treats 
       the gluon emission in the framework of the colour dipole 
       model~\cite{cdm} as an alternative to the parton showers in LEPTO. 
       In the selected phase space ARIADNE describes the data well
       with exception of the $R_{2+1}(\qq)$ distribution.
       The fitted $\almz$ value using fully simulated 
       ARIADNE events deviates by $-0.005$ from the quoted result obtained 
       with LEPTO.  
\end{itemize} 
The dependence of $\almz$ on the choice of the Monte Carlo model is estimated 
by the quadratic sum of all these uncertainties to be $+0.004$ and $-0.009$. 

Finally the theoretical uncertainty is investigated by changing various
input conditions for the NLO calculations:
\begin{itemize}
 \item The MRSH parton density function (PDF) used in MEPJET 
       is replaced by the MRSAp-201~\cite{MRSAp}, GRV92~\cite{GRV92}, 
       and the CTEQ-4A1~\cite{CTEQ} parameterizations. 
       They all use a similar   
       $\Lambda^{(4)}_{\scriptstyle\overline{\it{MS}}}$ value 
       of about 230~MeV. 
       The largest change in the fitted value of $\almz$ is $\pm 0.001$. 
       As a particular point of interest~\cite{Webber,chyla} 
       the correlation of the measured value of $\al$ in this analysis 
       with the value used while fitting the PDFs is investigated.
       This is done using the GRV92 PDFs which are available
       for various $\Lambda^{(4)}_{\scriptstyle\overline{\it{MS}}}$ 
       values~\cite{Vogt} in the range 0.15--0.40~GeV. 
       Although $\almz$ used in the PDFs changes by up to 0.017, 
       the fitted values only differ by $+0.003$ and $-0.002$. 
       This large reduction is due to the fact that the value of $\al$
       influences the scaling violation, whereas $R_{2+1}(\qq)$
       is directly proportional to the strong coupling constant.
       In table~\ref{errors} both errors were combined to one.
 \item The factorization scale $\mu^2_f$ and the renormalization
       scale $\mu^2_r$ are varied independently between $\qq/4$ and 
       $4\,\qq$. The effect on the measured value of $\almz$ is 
       $+0.003$ and $-0.007$. 
 \item The dependence on the chosen value of $y_{cut}$ was investigated 
       by varying $y_{cut}$ between 0.015 and 0.025 in the NLO calculation, 
       LEPTO simulation and for the data. Above 0.025 the number of (2+1) 
       jet events rapidly decreases but the results are still consistent 
       within the larger statistical error. Below 0.015 the regime of very 
       small jet masses opens up, making the fit procedure unstable. The 
       resulting error on $\almz$ is $+0.002$ and $-0.003$.
\end{itemize}
The total theoretical uncertainty ($+0.005$ and $-0.008$) on $\almz$  
is obtained summing these three errors in quadrature.  

The result of the analysis using the JADE algorithm, 
where the total systematic uncertainty is taken 
as the quadratic sum of all uncertainties 
discussed above, is: 
\[ 
\almz=\almeana\,.
\] 
The solid lines in figure~\ref{al3}b indicate the full range 
of $\alq$ obtained in this experiment using the JADE jet algorithm 
adding statistical and systematic errors in quadrature at 
the scale $\qq = \mzq$. 
They correspond to $\almz = 0.104$ and $0.126$. 

To obtain the systematic uncertainties on the four measured 
values of $\alq$ using the JADE algorithm the following method is applied. 
Each of the used scenarios, except the variation of the renormalization 
scale, results in measured values of $\alq$ for the four bins in 
$\qq$. From the values of the different cut scenarios the systematic 
experimental and theoretical uncertainties, as well as the model 
dependence in each bin, are determined in the same way as discussed 
above for the uncertainties on $\almz$. 
The total systematic uncertainty using the JADE algorithm 
is given by the quadratic sum of these three error contributions. 
The outer error bars of the $\al(\qq)$ points in figure~\ref{al3}b 
represent the total uncertainties, which are calculated 
as the quadratic sum of statistical errors and total systematic 
uncertainties (see table~\ref{values}). 

%
\begin{table}[t]
\begin{center}
\begin{tabular}{|c||c|c|c|c||c||c|}\hline
\va & \multicolumn{4}{c||}{ JADE } & E0  & P  \\\hline\hline
    $\qq$ [\gevq] & $\alq$  & stat  & sys & tot 
    & $\alq$ & $\alq$ \\\hline\hline
\va    54  & 0.184  & $\pm 0.017$ & $^{+\,0.035}_{-\,0.038}$ & 
                                   $^{+\,0.039}_{-\,0.042}$ 
     & 0.189 & 0.206 \\\hline
\va    209 & 0.166  & $\pm 0.009$ & $^{+\,0.028}_{-\,0.021}$ & 
                                    $^{+\,0.029}_{-\,0.023}$ 
     & 0.166 & 0.177 \\\hline
\va    440 & 0.151  & $\pm 0.010$ & $^{+\,0.018}_{-\,0.022}$ & 
                                    $^{+\,0.021}_{-\,0.024}$ 
     & 0.151 & 0.167 \\\hline
\va   1272 & 0.130  & $\pm 0.012$ & $^{+\,0.012}_{-\,0.020}$ & 
                                    $^{+\,0.018}_{-\,0.024}$ 
     & 0.127 & 0.128 \\\hline\hline
\va $\mzq$ & 0.117  & $\pm 0.003$ & $^{+\,0.009}_{-\,0.013}$ & 
                                    $^{+\,0.009}_{-\,0.013}$ 
     & 0.117 & 0.123 \\\hline
\end{tabular}
\caption[]{\it The values of $\alq$ together 
           with their statistical, systematic and total uncertainties 
           for the modified JADE algorithm at different values of $\qq$, 
           which are taken to be the  average in each considered $\qq$ 
           range (see table~1). In addition the extracted $\alq$ values 
           using the modified E0 and P algorithms are included.} 
\label{values}
\end{center}
\end{table}
%

To study the dependence of the determined $\almz$ value on different jet 
algorithms the analysis is redone for the central cut scenario using 
the E0 and P variants of the JADE jet algorithm. 
The obtained $\al$ values in the different bins of $\qq$ are given in 
table~\ref{values}. The extracted value of $\almz$ based on the E0 algorithm, 
$\almz = 0.117$, is, as expected from a theoretical point of view, 
very close to the value determined using the JADE algorithm, whereas for 
the P algorithm the fitted value is increased by $+0.006$. 
The final result of this analysis is therefore
\[ 
\almz=\almeanars\:,
\] 
where the last error reflects a possible sensitivity to the choice of the
jet algorithm used.

Another approach taken by the H1 Collaboration is to perform a quantitative 
pQCD analysis based on jet event rates differential~\cite{mweber} in the 
JADE jet resolution parameter. This method using a similar phase space region
leads to 
$\almz = 0.118 \pm 0.002 \,(stat) \, ^{+\,0.007}_{-\,0.008} \,(sys)
                                  \, ^{+\,0.007}_{-\,0.006} \,(theory)$, 
fully consistent in value and magnitude with the result of the present
analysis.

The combined result from studies of the hadronic final 
state in the experiments at LEP and SLC~\cite{PDG96} is 
$\almz = 0.122 \pm 0.007$. 
Here the error is totally dominated by 
theoretical and model dependencies associated with the same 
error sources as in the analysis presented in this paper. 
\section{Summary}
Jet production in neutral current deep inelastic 
{\it ep} scattering at HERA is studied using  
the modified JADE jet algorithm.
The strong coupling constant $\al$ is determined 
over a wide range of $\qq$ by evaluating jet event rates 
corrected to the parton level 
assuming the hadron-parton relation as given by 
the event generator LEPTO. 
The $\qq$ dependence of the determined $\al$ values 
is compatible with the theoretical 
prediction of the renormalization group equation. 
A fit of the QCD 
parameter~$\Lambda^{(4)}_{\scriptstyle\overline{\it{MS}}}$
leads to  
$ \almz = \almeana\, .$ 
The systematic error is dominated by the uncertainty in the hadronic 
energy scale, the renormalisation scale dependence, and the dependence 
on the Monte Carlo model. Adding statistical and systematic errors 
in quadrature yields $\almz= 0.117 ^{+\,0.009}_{-\,0.013}$ 
consistent with the world average $\almz = 0.118 \pm 0.003$~\cite{PDG96}. 
A dependence of the extracted $\almz$ value on variations of the JADE 
algorithm is found. 
The largest change observed is $+0.006$, leading to the final result:
\[ \almz = \almeanars\, .    \]
\section*{Acknowledgments}
We are grateful to the HERA machine group whose outstanding efforts have 
made and continue to make this experiment possible. We thank the engineers 
and technicians for their work in constructing and now maintaining the H1 
detector, our funding agencies for financial support, the DESY technical 
staff for continual assistance, and the DESY directorate for the hospitality 
which they extend to the non--DESY members of the collaboration. 
%


\clearpage

\end{document}